\documentclass[authoryear,12pt,3p]{jowarticle}
\usepackage{graphicx}
\usepackage{amsthm,amsmath}
\usepackage{amssymb}
\usepackage{amsfonts}
\usepackage{mathtools, eqparbox}
\usepackage{natbib}
\usepackage{setspace}
\usepackage[all]{xy}
\usepackage{enumitem}
\usepackage{titlesec}
\usepackage{mathrsfs}
\makeatletter
\renewcommand\section{\@startsection{section}{1}{\z@}{-3.25ex plus -1ex minus -.2ex}{1.5ex plus .2ex}{\normalsize\bf}}
\renewcommand\subsection{\@startsection{subsection}{2}{\z@}{-3.25ex plus -1ex minus -.2ex}{1.5ex plus .2ex}{\normalsize\bf}}
\renewcommand\subsubsection{\@startsection{subsubsection}{3}{\z@}{-3.25ex plus -1ex minus -.2ex}{1.5ex plus .2ex}{\normalsize\bf}}
\makeatother

\newtheorem{innercustomgeneric}{\customgenericname}
\providecommand{\customgenericname}{}
\newcommand{\newcustomtheorem}[2]{%
  \newenvironment{#1}[1]
  {%
   \renewcommand\customgenericname{#2}%
   \renewcommand\theinnercustomgeneric{##1}%
   \innercustomgeneric
  }
  {\endinnercustomgeneric}
}

\newtheorem{thm}{Theorem}
\newtheorem{cor}[thm]{Corollary}
\newtheorem{lem}[thm]{Lemma}
\newtheorem{prop}[thm]{Proposition}

\newcustomtheorem{prop*}{Proposition}
\newcustomtheorem{cor*}{Corollary}
\newcommand{\norm}[1]{\left\Vert#1\right\Vert}

\newcommand{\Reals}{\mathbb {R}}

\providecommand{\inner}[2]{\langle#1,#2\rangle}

\begin{document}
\begin{frontmatter}
\title{Why Be Regular?, Part II}

\author{Benjamin Feintzeig}\ead{bfeintze@uw.edu}
\address{Department of Philosophy \\ University of Washington}
\author{James Owen Weatherall}\ead{weatherj@uci.edu}
\address{Department of Logic and Philosophy of Science\\ University of California, Irvine}
\begin{abstract}We provide a novel perspective on ``regularity'' as a property of representations of the Weyl algebra.  In Part I, we critiqued a proposal by Halvorson [2004, ``Complementarity of representations in quantum mechanics", \textit{Studies in History and Philosophy of Modern Physics} \textbf{35}(1), pp. 45--56], who advocates for the use of the \emph{non-regular} ``position'' and ``momentum'' representations of the Weyl algebra.  Halvorson argues that the existence of these non-regular representations demonstrates that a quantum mechanical particle can have definite values for position or momentum, contrary to a widespread view.  In this sequel, we propose a justification for focusing on regular representations, \emph{pace} Halvorson, by drawing on algebraic methods.\end{abstract}
\end{frontmatter}
                                         % Activate to display a given date or no date
\doublespacing
\section{Introduction}
\label{sec:introduction}

In a provocative paper, \citet{HalvorsonBohr} argues that a quantum mechanical particle \emph{can} have a determinate position or momentum, contrary to a widespread view.  Following the algebraic approach to quantum theories, Halvorson starts by understanding the mathematical structure of the theory of a quantum mechanical particle to consist in an algebra capturing the canonical commutation relations (CCRs); Halvorson shows that if one employs a standard algebra known as the \emph{Weyl algebra}, then one can find a Hilbert space representation in which there exist determinate position states and another distinct Hilbert space representation in which there exist determinate momentum states.  Thus, Halvorson concludes that determinate position or momentum states appear within ordinary quantum theory.

These representations of the Weyl algebra containing determinate position and momentum states have been largely ignored because they lack a property called \emph{regularity}.  Halvorson believes the regularity assumption is suspect, and advocates for exploring the consequences of giving it up. However, we believe that there is good reason to restrict attention to regular representations.  The purpose of this paper is thus to provide a justification for regularity, \emph{pace} Halvorson.

The present paper is the second in a two part series.\footnote{Although this paper appears as Part II in a series, it is self-contained and may be read independently of Part I.  We set up the (part of the) problem we discuss here in section \ref{sec:prelim}.}  In Part I \citep{FLRW}, we offered a critical analysis of Halvorson's proposal for how and why to relax regularity, on several ways of understanding what he may have intended.  We ultimately concluded that the most attractive reading of his proposal is one on which the position and momentum representations themselves play a relatively minor role, and instead one focuses on the relationship between regular and non-regular states on the abstract Weyl algebra.  We concluded that paper by arguing that the appeal of this proposal turns on the choice of the Weyl algebra, as opposed to other possible C*-algebras, to represent the physical magnitudes associated with a particle---a choice that Halvorson does not motivate or defend.

In the present paper we will take up the question of how one might justify a choice of algebra and argue that the Weyl algebra is an injudicious choice.  We will show that a more judicious choice of algebra provides independent grounds for taking only regular representations to be physically significant.  The states of non-regular representations (e.g., determinate position and momentum states), meanwhile, are best understood as idealized limiting cases, in some ways ``unphysical'' but nonetheless approximating physically realizable states.  We will support these claims by proving a number of small propositions.

In order to prove our results, we will need to make some continuity assumptions about states, and so our discussion will touch upon ideas about continuity of states that Halvorson explores in an earlier paper (\citeyear{HalvorsonTeller}), presented in response to \citet{Teller}.  Halvorson's strategy in the \citeyear{HalvorsonTeller} paper, like that in the later (\citeyear{HalvorsonBohr}) paper, is to find quantum states with determinate positions in places that have been overlooked in the existing mathematical formalism of quantum mechanics.  However, instead of appealing to non-regular representations, the earlier (\citeyear{HalvorsonTeller}) paper identifies determinate position states as \emph{non-normal} states, which fail to satisfy certain continuity requirements, on an algebra of position observables.

Halvorson does not explicitly discuss any connection between (non-)normality and (non-) regularity, even though these conditions appear at crucial junctures in distinct arguments for the same conclusion concerning the significance of determinate position states.  However, the condition of normality has been discussed elsewhere by \citet{RuetscheNormal}, who argues that only normal states respect ``lawlike'' relations between physical magnitudes.  Though even Ruetsche admits that this abstract justification for normality is not quite enough:
\begin{quote}
My unsurprising conclusion is that we can't decide which states are physical in isolation from considering what lawlike relationships we care about, and what magnitudes those relationships involve \citep[p. 115]{RuetscheNormal}
\end{quote}
According to Ruetsche, appeal to conditions like normality must be supplemented by context-specific interpretive work to determine the physical significance of algebraic tools.

In this paper, we attempt to do this supplemental work for the quantum theory of a single free particle.  We provide a strategy for determining the lawlike relationships we care about, and we show that following this strategy leads to a formalism in which regularity and normality line up.  Under these circumstances, Ruetsche's defense of normality becomes a defense of regularity.  Our contribution is to show how careful attention to the way we use the mathematical tools of quantum theories to represent physical situations establishes a connection between normality and regularity, which we use to justify the latter condition.

The plan of the paper is as follows.  In \S\ref{sec:prelim}, we describe the relevant background for the current paper and set up the problem by sketching Halvorson's view on non-regular representations and reviewing, in somewhat more technical detail than we offer in Part I, our argument that the appeal of Halvorson's proposal turns on his choice of the Weyl algebra.   Our main argument takes place in the following two sections, in which we use this algebraic perspective to justify using an alternative algebra that essentially admits only regular representations (\S\ref{sec:justification}) and show a precise sense in which states in non-regular representations can be recovered, albeit as idealizations from states in regular representations (\S\ref{sec:approximation}).  We conclude with some general discussion of the interpretation of quantum theories in \S\ref{sec:conclusion}.  Background technical material and proofs of propositions appear in two appendices.

\section{Background}
\label{sec:prelim}

\citet{Ruetsche} distinguishes broadly between two approaches to interpreting quantum theories: that of the \emph{Hilbert space conservative} and that of the \emph{algebraic imperialist}.  Roughly, the Hilbert space conservative puts the physical content of a quantum theory in some concrete Hilbert space representation of the canonical commutation relations, whereas the algebraic imperialist puts the physical content in an abstract algebra of physical magnitudes.  In Part I of this series, we considered Halvorson's use of non-regular representations from the perspective of a Hilbert space conservative, who would attempt to use the states and magnitudes in non-regular representations as in the Schr\"{o}dinger representation.  We found such an interpretation wanting due to a number of conceptual and mathematical obstacles.

In the penultimate section of Part I, we sketched a different approach to understanding Halvorson's proposal, based on an algebraic imperialist perspective.\footnote{The ``algebraic imperialist" approach considered here differs somewhat from that described by \citet{Ruetsche}, which is tied to what she calls ``pristine interpretation''.  For the purposes of this paper, an algebraic imperialist approach is one that uses broadly algebraic tools in the abstract without focusing on particular representations, but is not necessarily pristine.}  Indeed, as we observed there, such an approach might come closer to Halvorson's intended interpretation as expressed in the following remark:
\begin{quote}
The abstract Weyl algebra carries the full empirical content of the quantum theory of a single particle.  In particular, the Weyl algebra has enough observables to describe any physical measurement procedure and enough states to describe any laboratory preparation.  A representation does not make any further empirical predictions; indeed, it adds \emph{nothing} in the way of empirical content. \citep[][p. 55]{HalvorsonBohr}
\end{quote}
This paper will investigate the plausibility of this algebraic imperialist interpretation of non-regular representations. In the present section, we will revisit---in somewhat more technical detail---the arguments sketched in Part I concerning what Halvorson's proposal might be, in order to pinpoint the issues on which its attractiveness turns.

Recall that to construct the theory of a single quantum mechanical particle moving in one dimension,\footnote{All statements of this paper generalize to a system with phase space $\mathbb{R}^{2n}$.  For example, this includes any finite number of free particles moving in any finite number of spatial dimensions represented by $\mathbb{R}$.} one finds a Hilbert space $\mathscr{H}$ along with a pair of self-adjoint operators, $Q$ and $P$, satisfying the canonical commutation relations:
\[[Q,P] = iI\]
Here, $I$ is the identity operator, and we work in units so that Planck's constant takes the numerical value $\hbar = 1$.  Although the above relation for $Q$ and $P$ is thought to capture much of the physical content of the quantum theory, since $Q$ and $P$ are in general unbounded, it is technically more convenient to work instead with certain bounded functions of them.  Following \citet{Weyl}, it is standard to consider formal exponentiations $U_a = e^{iaQ}$ and $V_b = e^{ibP}$ for $a,b\in\mathbb{R}$.  One hopes to recover the operators $Q$ and $P$ by ``differentiation'' via Stone's theorem.  This is always possible if appropriate continuity conditions are satisfied \citep[see, e.g.,][p. 264]{ReSi80}.  For these bounded operators, the commutation relations are expressed as:
\begin{align*}
&&U_aV_b = e^{-iab}V_bU_a && \mathrlap{a,b\in\mathbb{R}}.
\end{align*}

To consider Hilbert space representations of this bounded form of the commutation relations, we form the smallest C*-algebra\footnote{For a general introduction to the algebraic tools used in this paper, see \ref{app:background}.  For further mathematical background, see \citet{Sa71,Di77,Ta79,KaRi97}.  For applications to quantum theory, see \citet{Em72,Ha92,La17}.  For philosophical introductions, see \citet{Ha06,Ruetsche}.} containing the operators $U_a$ and $V_b$ \citep[See][]{Pe90}.  To do so, consider the formal products $W_{a,b} = e^{iab/2}U_aV_b$.  (One can always recover $U_a = W_{a,0}$ and $V_b = W_{0,b}$ from such operators.)  Now the commutation relations become:
\begin{align*}
&&W_{a,b}W_{a',b'} = e^{\frac{i}{2}(a'\cdot b - b'\cdot a)}W_{a+a',b+b'}&& \mathrlap{a,a',b,b'\in\mathbb{R}}
\end{align*}
Consider the vector space generated by formal linear combinations of the Weyl operators $W_{a,b}$ for all $a,b\in\mathbb{R}$ with the above multiplication operation.  One can define an involution on these operators by $(W_{a,b})^* := W_{-a,-b}$.  Then there is a unique norm one can define on the resulting involutive, associative algebra (the so-called minimal regular norm) so that, when completed, the resulting collection of operators forms a C*-algebra \citep{MaSiTeVe74,BiHoRi04a}.  We call the resulting C*-algebra the \emph{Weyl algebra}\footnote{More precisely, this is the Weyl algebra \emph{over $\mathbb{R}^2$} since one has an operator $W_{a,b}$ for each point $(a,b)$ in the phase space $\mathbb{R}^2$.  More on the relationship between Weyl operators and classical phase space will be discussed in \S\ref{sec:justification}.} and denote it by $\mathcal{W}$.

We call a representation $(\pi,\mathscr{H})$ of $\mathcal{W}$ \emph{regular} just in case for all $\varphi,\psi\in\mathscr{H}$, the maps
\begin{align*}
a\mapsto \inner{\varphi}{\pi(W_{a,0})\psi} && \text{ and } && b\mapsto \inner{\varphi}{\pi(W_{0,b})\psi}
\end{align*}
are continuous.  Regular representations are special because they allow one to simultaneously define, via Stone's theorem, self-adjoint operators $Q$ and $P$ generating the one parameter families $W_{a,0}$ and $W_{0,b}$, which we interpret as representing the physical magnitudes of position and momentum, respectively.

It is well known that in any regular representation of the Weyl algebra the corresponding position operator $Q$ fails to have any eigenvectors \citep{HalvorsonBohr}.  For example, this statement holds in what is known as the \emph{Schr\"{o}dinger representation}, which is the representation one standardly considers in textbook quantum theory.  If one believes that only eigenvectors can have a determinate value for a physical quantity, then one arrives at the widespread view that in no quantum state does a single particle have a determinate position, as described in \citet{Teller}.

However, \citet{HalvorsonBohr} argues against this standard interpretation.  Halvorson explicitly constructs two non-regular representations of the Weyl algebra \citep{FLRW}: one that we refer to as the \emph{position representation} in which there exist vectors representing states with determinate positions, and another that we refer to as the \emph{momentum representation} in which there exist vectors representing states with determinate momenta \citep[see also][]{BeMaPeSi74,FaVeWe74}.  Furthermore, Halvorson is able to capture what he takes to be a feature of Bohr's complementarity principle by showing that in any representation in which the position operator $Q$ exists and has an eigenvector, the momentum operator $P$ does not exist, and vice versa.  This implies, in other words, that determinate position states can only appear as eigenvectors of $Q$ in a representation of the Weyl algebra if that representation fails to be regular; and similarly for momentum states.

So, according to Halvorson, there are interpretive benefits to considering non-regular representations of the Weyl algebra.  Non-regular representations allow one to recover the existence of states with determinate quantities and a notion of position-momentum complementarity.  Of course, one could attempt to defend the standard interpretation of quantum states as not possessing determinate position or momentum values by taking on the assumption of regularity and appealing to the well-known Stone-von Neumann theorem,\footnote{See, e.g., \citet{Su99,ClHa01,Ruetsche}.} which establishes that all irreducible regular representations are equivalent to the Schr\"{o}dinger representation.  But Halvorson rejects this argument, asserting that ``\emph{the regularity assumption begs the question against position-momentum complementarity}" \citep[p. 49]{HalvorsonBohr}.  For this defense of the standard interpretation to be convincing, one requires independent justification for the regularity assumption, which we will attempt to give in what follows.

However, before turning to our main argument, it is worth a further remark to try to make sense of Halvorson's view.  What reason could Halvorson possibly have to think of states in inequivalent representations of the Weyl algebra as physically significant?  In Part I, we suggest reading Halvorson as implicitly appealing to a well-known result of \citet{Fe60}\footnote{See also \citet{Kronz+Lupher} for a philosophical discussion.} to justify his use of states on the Weyl algebra that do not appear in the ordinary Schr\"{o}dinger representation.  Fell's theorem establishes the relationship between states in different irreducible representations of a C*-algebra.  Namely, it follows from Fell's theorem that states in the Schr\"odinger, position, and momentum representations can all be approximated by one another in a particular topology determined by the Weyl algebra.

Here is slightly more detail.  The weak* topology on the state space of the Weyl algebra\footnote{The weak* topology on the state space of the Weyl algebra is just one instance of the more general definition of the weak* topology on the dual space to a C*-algebra.  See \ref{app:background}.} is given by the following condition for convergence (pointwise on elements of $\mathcal{W}$): a net of states $\omega_\alpha$ converges to a state $\omega$ in the weak* topology iff for each $A\in\mathcal{W}$, $\omega_{\alpha}(A)\rightarrow\omega(A)$.  Fell's theorem establishes that the folium of any faithful representation of a C*-algebra is dense in the folium of any other faithful representation in the weak* topology on the state space.  In other words, every state with a density operator representative in one faithful representation can be approximated as a limit in the weak* topology of states with density operator representatives in another faithful representation.  Now it suffices to notice that the Schr\"odinger, position, and momentum representations are all faithful,\footnote{In fact, since the Weyl algebra is \emph{simple}, all of its nontrivial representations are faithful.} which establishes that the corresponding states in each folium can all be used to approximate one another.

The above sketch of an argument provides one sense in which the Schr\"odinger, position, and momentum representations of the Weyl algebra might be understood to be on a par, at least with respect to the collection of states they appear to deem physically possible.  This is at least one way of making precise Halvorson's claim that the different representations have the same empirical content.  It also provides a sense in which one might freely choose states from different representations, depending on context.  We take it that this is probably the most charitable interpretation of Halvorson's proposal; even if it is not the interpretation Halvorson intended, we believe it is one worth considering. 

Still, some care is needed.  The interpretation we have just given on behalf of Halvorson relies on the assumption that the weak* topology on the state space of $\mathcal{W}$ provides a physically relevant notion of approximation for states.  But does it? Or, put differently, does the Weyl algebra, which determines this weak* topology, have the physical significance Halvorson attributes to it?  In what follows, we challenge both of these implicit assumptions.

In the following section, we will point out that there may be reasons to use a different algebra of quantities for our quantum theory, and hence a different state space.  But the weak* topology is determined by our choice of algebra and in general will differ if we use a different algebra of quantities.  So even if the weak* topology on a state space captures a natural notion of approximation, it is not at all clear that the weak* topology on the state space \emph{of the Weyl algebra in particular} has the physical significance attributed to it in the above Halvorsonian argument.  And if the weak* topology does not have an appropriate physical significance, then the outlined motivation for considering states in non-regular representations diminishes in strength.

But suppose we are willing to stick with our algebraic perspective as we follow others in considering alternative algebras for our quantum theory.  We show in \S\ref{sec:justification} that if one imposes at least a necessary condition on the algebra of quantities we choose for our quantum theory, then it follows that any physical representation of this algebra must be regular.  However, we go on to show in \S\ref{sec:approximation} that states in non-regular representations can still be recovered as idealizations or approximations even if states in regular representations are privileged, thus recapturing the content of the argument based on Fell's theorem in a new light.

\section{A Justification for Regularity}
\label{sec:justification}

Although many have noted reasons against employing the Weyl algebra \citep[see, e.g.,][]{FaVe74,BuGr08,Fe18}, the majority of the philosophical community concerned with algebraic methods in quantum theories, including Halvorson, still exclusively uses this C*-algebra.  In this section, we argue that (i) the possibility of choosing an alternative algebra undermines the significance of Fell's theorem spelled out in the previous section, and (ii) if one imposes a necessary condition on the choice of a new quantum algebra proposed by \citet{Fe17b}, then one rules out non-regular representations altogether.

\subsection{Quantization and the choice of algebra}

Why might we consider alternative algebras of magnitudes for a quantum theory?  One reason is that the Weyl algebra arises through the quantization of a particular classical algebra of quantities; but there are other algebras of classical quantities that may serve as a natural starting point for quantization, at least for some purposes.\footnote{This is not the only reason to consider alternative quantum algebras.  For example, \citet{BuGr08} consider alternative algebras because the Weyl algebra fails to allow for enough dynamical automorphisms \citep[see also ][]{FaVe74}, \citet{La90,La90b} considers alternative algebras in order to accommodate classical symmetries, and \citet{GrNe09} consider alternative algebras for the express purpose of ruling out non-regular representations.}

The Weyl algebra is the collection of physical quantities obtained by quantizing the following collection of classical physical quantities.\footnote{For more details, see \citet{BiHoRi04a,BiHoRi04b}.}  The C*-algebra $AP(\mathbb{R}^{2})$ of \emph{almost periodic functions} on $\mathbb{R}^{2}$ is the smallest C*-algebra containing the functions $w_{a,b}:\mathbb{R}^{2}\rightarrow\mathbb{C}$ defined for each pair $a,b\in\mathbb{R}$ by
\begin{align*}
&& w_{a,b}(q,p) = e^{a\cdot q + b\cdot p}&& \mathrlap{q,p\in\mathbb{R}}
\end{align*}
The operations on $AP(\mathbb{R}^{2})$ are given by pointwise addition, multiplication, scalar multiplication, and complex conjugation with the standard supremum norm.  This is an algebra of classical quantities on the classical phase space $\mathbb{R}^{2}$, but we can quantize it by using the following linear, *-preserving quantization map.  Define $\mathcal{Q}: AP(\mathbb{R}^{2})\rightarrow \mathcal{W}$ as the unique linear, continuous extension of
\begin{align*}
&&\mathcal{Q}(w_{a,b}) := W_{a,b} && \mathrlap{a,b\in\mathbb{R}}
\end{align*}
In fact, \citet{BiHoRi04a,BiHoRi04b} show that this quantization map defines what is known as a strict deformation quantization of $AP(\mathbb{R}^2)$.

Once we see how the Weyl algebra is constructed from analogous classical quantities, it becomes unclear whether there is any sense in which this algebra is privileged.  After all, it is not obvious that the classical algebra of almost periodic functions, which we quantize to arrive at the Weyl algebra, has any particular \emph{physical} significance.  It is certainly technical convenient for expressing the CCRs in bounded form and for developing the rich theory of Fourier analysis in both the classical and quantum context \citep[see][]{We84}.  But these instrumental benefits do not immediately give rise to interpretive values.

And indeed, many have considered quantizing different classical algebras.  For example, other well known quantization procedures begin instead with the algebra $C_0(\mathbb{R}^2)$ of \emph{continuous functions vanishing at infinity} \citep[see, e.g.,][]{La98b,La06,La17}.  This algebra is a nice choice because its state space consists of precisely the Borel probability measures on $\mathbb{R}^2$, a seemingly natural choice for the classical theory because these may be understood as probability distributions over instantaneous classical determinate states.  However, it can be somewhat hard to compare quantization maps on $C_0(\mathbb{R}^2)$ with those on $AP(\mathbb{R}^2)$ because these domains are disjoint.  So in what follows we will also consider the W*-algebra $B(\mathbb{R}^2)$ of bounded measurable functions,\footnote{Here, we consider functions measurable with respect to the $\sigma$-algebra of universally Radon measurable sets.  The universally Radon measurable sets include any set that is measurable with respect to the completion of every bounded regular Borel measure.  So at least every Borel measurable function belongs to $B(\mathbb{R}^2)$.  For definitions and background mathematical theory, see \citet{Fr03}.} which is an algebra of classical quantities containing both $AP(\mathbb{R}^2)$ and $C_0(\mathbb{R}^2)$ as subalgebras.  Thus, we have the following inclusion relations:
\begin{center}
\leavevmode
\xymatrix{
C_0(\mathbb{R}^{2}) \ar@{}[dr]|-*[@]{\subset}& \\
& B(\mathbb{R}^{2}) \\
AP(\mathbb{R}^{2}) \ar@{}[ur]|-*[@]{\subset} &
}
\end{center}
It is known that $B(\mathbb{R}^2)$ is the bidual of $C_0(\mathbb{R}^2)$.\footnote{See \ref{app:background} or \citet{Fe17c} for background on W*-algebras and biduals.  See Prop. 437I of \citet{Fr03} for the stated result.}   This implies that $B(\mathbb{R}^2)$ contains all and only the weak limits of quantities in $C_0(\mathbb{R}^2)$, which can be understood as approximate or idealized physical quantities \citep{Fe17c}. Furthermore, it follows that the normal state space of $B(\mathbb{R}^2)$ is identical with the state space of $C_0(\mathbb{R}^2)$ and thus also consists of all Borel probability measures on $\mathbb{R}^2$.  (Recall that a \emph{normal} state on a W*-algebra is one that belongs to the predual of that algebra, or, equivalently, one that is continuous in the weak* topology on that algebra.  See \ref{app:background}.)  So $B(\mathbb{R}^2)$ is, in a sense, another natural choice for an algebra of classical quantities.

Now, as soon as one starts considering alternative classical algebras, one also considers different quantum algebras.  For example, \citet{La98b} quantizes $C_0(\mathbb{R}^2)$ to obtain the compact operators on a Hilbert space (or more generally, a twisted group algebra).  And \citet{BuGr08} quantize a different classical algebra to obtain what they call the resolvent algebra.  These algebras at least \emph{prima facie} compete with the Weyl algebra for the title of algebra of quantum magnitudes.

Considering different quantum algebras undermines the use of Fell's theorem in the previous section.  Different algebras have different state spaces, and even if we can identify \emph{some} of the states on different algebras with each other, the weak* topologies on these state spaces will differ.  So if we forgo the use of the Weyl algebra in favor of some other physically significant collection of quantities, it is not clear that Fell's theorem can be used to argue that states in regular and non-regular representations can be used to approximate one another in a physically meaningful way.  In fact, we will show in \S\ref{sec:approximation} that on an alternative (and from our perspective, preferred) choice of algebra, one can show that the approximation relation holds only in one direction and not in the other, thereby privileging some states over others.

In fact, it is not even clear that different algebras we might choose will allow for non-regular representations.  For example, \citet{La90b} provides a construction procedure for an algebra that does not admit any non-regular representations, and \citet{GrNe09} attempt to generalize this procedure to systems with infinitely many degrees of freedom.  So it is not at all clear from an algebraic perspective that non-regular representations of the Weyl algebra are on a par with regular representations, because other choices for the algebra of quantum magnitudes rule out non-regular representations altogether.

All this leads to what we previously said was the central question: what algebra should be choose?  Next, we propose one possible perspective on what might guide this choice.  We will argue that the Weyl algebra is not a natural or compelling choice.  We show that if one is guided instead by the constraints of the classical theory one is quantizing in a certain way when one chooses an alternative algebra of magnitudes for a quantum theory, then all representations that ``preserve algebraic structure'' (in a certain sense) must be regular.

\subsection{Quantization leads to regular representations}

We now turn to our argument that regular states have a special status after all, and that the Weyl algebra is an injudicious choice for representing physical magnitudes.  We will not endorse a particular alternative to the Weyl algebra; instead, we will show that for any algebra arising as the codomain of a quantization map satisfying a certain constraint, all of the representations that ``preserve algebraic structure" (in a certain sense) are regular.  We do not claim that the argument we offer is the only, or even the best, justification for working with a different algebra from the Weyl algebra.  But we do claim that it is well-motivated, and that it provides a justification for considering only regular representations. 

The constraint we will impose on our quantization map is that it be appropriately \emph{continuous}, as argued for in \citet{Fe17b}.\footnote{We acknowledge that there is a worry, here, that our proposal looks trivial: the regularity condition amounts to requiring certain maps to be continuous; we justify it by requiring some other maps to be continuous.  Why should anyone willing to reject regularity accept another continuity condition instead?  Likewise, below, we argue that normal states have a special status; these, too, are just states that are continuous in a certain topology.  But this objection is to miss the point of our argument, which is that continuous maps preserve topological structure and thus, if we have reason to think that a particular topology captures physically salient information, maps that are continuous with respect to that topology respect that structure.  We thus get to regularity (and normality) by starting with a topology whose significance we believe is manifest, and then insisting on preserving that structure through the quantization procedure.}  We will not attempt to justify this constraint; we merely note that Feintzeig argues that this continuity constraint captures our desire to have a notion of approximation on our quantum algebra that we can naturally interpret as corresponding to a notion of approximation on our classical algebra.  One can understand all of what follows in this section to be provisional on the acceptance of the conclusion of Feintzeig's previous argument.  Requiring continuity of a quantization map is not novel; the key idea, however, that we will take from Feintzeig is his suggestion about \emph{with respect to which topologies} our quantization map ought to be continuous.

First, some setup.  We will consider quantization maps $\mathcal{Q}:\mathfrak{C}\rightarrow\mathfrak{R}$, where $\mathfrak{C}\subseteq B(\mathbb{R}^2)$ is some algebra of classical quantities.  We will require that $AP(\mathbb{R}^2)\subseteq \mathfrak{C}$ so that our quantum algebra contains the almost periodic functions, which implies that we at least have a surrogate $\mathcal{Q}[AP(\mathbb{R}^2)]$ for the Weyl algebra.  This is necessary for us to even be able to state the condition of regularity with respect to the Weyl unitaries, as they arise from the quantized versions of almost periodic functions.  In this setting, we will say that a representation $(\pi,\mathscr{H})$ of the quantum algebra $\mathfrak{R}$ is \emph{regular} if for every $\varphi,\psi\in\mathscr{H}$, the maps
\begin{align*}
a\mapsto\inner{\varphi}{\pi\circ\mathcal{Q}(w_{a,0})\psi} && \text{ and } && b\mapsto\inner{\varphi}{\pi\circ\mathcal{Q}(w_{0,b})\psi}
\end{align*}
are continuous.

Following Feintzeig, we are interested in preserving the notion of approximation of quantities by their expectation values on physical states, which is encoded in the classical theory by the weak* topology on $B(\mathbb{R}^2)$.  Recall this weak* topology on $B(\mathbb{R}^2)$ agrees with the topology of pointwise convergence on bounded sequences of functions, which carries a natural physical interpretation.  Moreover, the weak* topology encodes approximation relations with respect to the expectation values of normal states, which \citet{RuetscheNormal} has argued are physically significant on general grounds, and which appear to be significant in this context because normal states on $B(\mathbb{R}^2)$ are precisely the Borel probability measures on the phase space $\mathbb{R}^2$ (more on this in the next subsection).  Thus, the weak* topology on $B(\mathbb{R}^2)$ provides a physically significant notion of approximation in the classical theory.  

We will also suppose that the quantum algebra $\mathfrak{R}$ is a W*-algebra with physically significant states in its normal state space.\footnote{One might worry that this assumption---that our quantum algebra is a W*-algebra---is too strong.  However, we need not worry because if our quantum algebra is a mere C*-algebra $\mathfrak{A}$, then we can always form a W*-algebra by taking its bidual $\mathfrak{A}^{**}$ (See \ref{app:background}).  We deal with this scenario in Cor. \ref{cor:c*}.}  In this case, the weak* topology on $\mathfrak{R}$ also encodes a physically relevant notion of approximation as long as the normal states of $\mathfrak{R}$ are indeed the physically significant ones.  We discuss the significance of normal states further below, but for now let us take as an assumption that the weak* topologies on $B(\mathbb{R}^2)$ and $\mathfrak{R}$ have the physical interpretation just outlined---see \citet{Fe17b} for more detail and motivation.

With our background in place, we now express the constraint we will impose: we require that the quantization map $\mathcal{Q}$ be continuous with respect to (a) the subspace topology on $\mathfrak{C}$ induced by the weak* topology of $B(\mathbb{R}^2)$ and (b) the weak* topology of $\mathfrak{R}$.  For shorthand, we will simply say that such a quantization map is \emph{weakly continuous}.\footnote{Note that a weakly continuous $\mathcal{Q}$ in our sense may fail to be continuous in the weak topology on $\mathfrak{C}$ and the weak topology on $\mathfrak{R}$, or in the weak topology on $\mathfrak{C}$ and the weak* topology on $\mathfrak{R}$.  The use of the weak* topology on $B(\mathbb{R}^2)$ is essential, and this is precisely where we bring in context-dependent interpretive assumptions that make our algebraic approach unpristine, in the terminology of \citet{Ruetsche}.}  Again, our conclusions in what follows should be understood as provisional on the acceptance of this condition.

Finally, since $\mathfrak{R}$ is assumed to be a W*-algebra, we consider what \citet{Sa71} calls w*-representations.  A representation $(\pi,\mathscr{H})$ of a W* algebra $\mathfrak{R}$ a \emph{w*-representation} just in case $\pi$ is continuous in the weak* topology on $\mathfrak{R}$ and the weak operator topology on $\mathcal{B}(\mathscr{H})$.  A w*-representation can be understood to preserve the W*-algebraic structure of $\mathfrak{R}$ by including only normal states in its folium.  Hence, restricting attention to w*-representations focuses on representations that ``preserve algebraic structure'' in the appropriate sense.  More will be said on this point, but we save discussion for what follows and now present a first result.\footnote{Proofs of all propositions are contained in \ref{app:proofs}.}
\begin{prop}
\label{prop:representation}
Let $\mathfrak{C}$ be a C*-algebra with $AP(\mathbb{R}^2)\subseteq \mathfrak{C}\subseteq B(\mathbb{R}^{2})$.  Let $\mathcal{Q}:\mathfrak{C}\rightarrow\mathfrak{R}$ be a linear, *-preserving map into a W*-algebra $\mathfrak{R}$.  Suppose further that $\mathcal{Q}$ is weakly continuous.  Then every w*-representation of $\mathfrak{R}$ is regular.
\end{prop}

\noindent The intended interpretation of Prop. \ref{prop:representation} is that if we restrict attention to appropriate algebras, then all representations that ``preserve algebraic structure" are regular.

We have expressed this result in terms of w*-representations.  Some readers may balk at this, or may find our justification for considering only w*-representations---that they ``preserve the W* algebraic structure of $\mathfrak{R}$''---obscure. But there is another way of thinking about what has been proven here.  One can justify restricting attention to w*-representations by noting that the GNS representation of any normal state on a W*-algebra is automatically a w*-representation.  This means that we are restricting attention to precisely those representations that arise as the ``home'' representations of the normal states.\footnote{For background on the GNS construction, see \ref{app:background} and \citet{KaRi97}.}  We state this as a corollary to the previous proposition.

\begin{cor}
\label{prop:normalrep}
Let $\mathfrak{C}$ be a C*-algebra with $AP(\mathbb{R}^2)\subseteq \mathfrak{C}\subseteq B(\mathbb{R}^{2})$.  Let $\mathcal{Q}:\mathfrak{C}\rightarrow\mathfrak{R}$ be a linear, *-preserving map into a W*-algebra $\mathfrak{R}$.  Suppose further that $\mathcal{Q}$ is weakly continuous.  Then for any normal state $\omega$ on $\mathfrak{R}$, the GNS representation $(\pi_\omega,\mathscr{H}_\omega)$ for $\omega$ is regular.
\end{cor}

\noindent The intended interpretation of Cor. \ref{prop:normalrep} is that if we restrict attention to the GNS representations of normal states on appropriate algebras, then all such representations are regular.

This result provides some justification for restricting attention to w*-representations. But it also raises a second worry.  Why should we restrict attention to GNS representations of normal states?  Indeed, why focus on normal states at all?  We have already noted that \citet{RuetscheNormal} has argued that normal states are the physically significant ones, but this view is controversial.  Perhaps most saliently, \citet{HalvorsonTeller} explicitly rejects any restriction to normal states in his response to \citet{Teller}, and so it is not clear why he should be moved by a justification for regularity that begins by focusing on normal states.\footnote{Recall that Halvorson's discussion in (\citeyear{HalvorsonTeller}) is somewhat different from his discussion in (\citeyear{HalvorsonBohr}).}   To deal with this objection, it is worth a digression to understand Halvorson's rejection of the standard restriction to normal states and the role it plays in his advocacy for determinate position states.  This will additionally help us clarify the (unpristine) algebraic perspective taken in this paper and our methodology for interpreting quantum theories.

\subsection{A digression concerning normal states}

\citet{HalvorsonTeller} considers states on the W*-algebra $L^\infty(\mathbb{R})$ of equivalence classes of bounded measurable functions with respect to the Lebesgue measure on $\mathbb{R}$.\footnote{Really, Halvorson restricts attention to the projection lattice of $L^\infty(\mathbb{R})$, which consists of equivalence classes of Borel measurable sets that differ by a set of Lebesgue measure zero.}  The algebra $L^\infty(\mathbb{R})$ can be embedded in the collection $\mathcal{B}(\mathscr{H})$ of all bounded operators on the Hilbert space $\mathscr{H} = L^2(\mathbb{R})$ of the Schr\"{o}dinger representation, acting on vectors by multiplication.  Seen this way, $L^\infty(\mathbb{R})$ is the W*-algebra generated by the standard position operator for a quantum particle (multiplication by $x$ on $L^2(\mathbb{R})$).  Halvorson argues that definite position states are physically significant because they appear as states, albeit \emph{non-normal} states, on this particular algebra $L^\infty(\mathbb{R})$.  Thus, Halvorson's earlier work comes to the same conclusion as his later (\citeyear{HalvorsonBohr}) paper---that definite position states are physically significant---but this time by rejecting normality rather than regularity.

\citet{RuetscheNormal} responds to Halvorson by arguing that \emph{if} we have chosen an appropriate algebra, \emph{then} we have good reason to restrict attention to normal states.  This is because, according to Ruetsche, normal states are the only states that obey the ``lawlike" relationships between physical magnitudes encoded in the algebra that they are states on.  If Ruetsche is correct, then one is justified in restricting attention to normal states insofar as one begins with an algebra that does indeed capture the appropriate ``lawlike" relationships between physical magnitudes.  

We basically agree with Ruetsche on this point.  In fact, we think Ruetsche's arguments are particularly persuasive against Halvorson because he equivocates on which quantum algebra to consider---the ``position algebra" $L^\infty(\mathbb{R})$ in (\citeyear{HalvorsonTeller}) and the Weyl algebra $\mathcal{W}$ in (\citeyear{HalvorsonBohr})---and thus, by Ruetsche's lights, he equivocates on which lawlike relationships he desires to capture.  Halvorson provides no independent justification for the use of these algebras in the contexts considered.  And the difference matters! Since $\mathcal{W}$ is not a W*-algebra one cannot distinguish normal states, while on $L^\infty(\mathbb{R})$ one cannot state the regularity condition because one does not have momentum unitaries.  From our perspective, appeal to conditions like regularity and normality cannot help us decide which states are physically significant until we have agreed on an appropriate algebra of physical quantities.  So it seems Ruetsche is correct that discussion of which states are physically significant should go beyond consideration of the conditions of regularity and normality in the abstract.

But Ruetsche does not give any further guidance on how one might tell whether one has captured the ``right'' lawlike relations with one's choice of algebra.  In this subsection, we will attempt to justify restricting attention to normal states in the current context by considering the significance of normal states in the classical theory, and then noticing that a quantization map allows us to draw an analogy between the classical and quantum case \citep[see also][]{Fe17b}. This justification for focusing on normal states is very closely related to the arguments of the previous subsection; in both cases, we use our background knowledge of the relevant classical theory to determine a physically salient notion of continuity that we then demand be preserved in the quantum context.  Thus we see that regularity and normality are not only both justified, but that they are justified by closely related considerations.

Before we can proceed with this argument, however, we must note that \citet{HalvorsonTeller} appears to disagree with our description of \emph{classical} theories above.  We now pause to justify our use of the algebra $B(\mathbb{R}^2)$ in the classical case.  Our discussion of normal states in classical theories forms the basis for the argument we give for restricting attention to normal states in quantum theories.  This also serves to illustrate the theme of this paper that careful attention to how we use algebraic tools is required in order to interpret physical theories.

Halvorson's description of classical theories differs from ours in that he argues the algebra $L^\infty(\mathbb{R})$ is also an appropriate algebra for representing the position magnitudes of a \emph{classical} particle, as opposed to the algebra of bounded measurable functions we considered above.  If Halvorson is correct, this would undermine the assumptions of Prop. \ref{prop:representation} and Cor. \ref{prop:normalrep} in which we use the classical algebra of bounded measurable functions without identifying functions equal almost everywhere with respect to Lebesgue measure.\footnote{\label{fn:configurationphase}Although Halvorson considers $L^\infty(\Reals)$ because he is thinking about quantities defined on a position configuration space $\Reals$, in this paper we work in the more general setting of the phase space $\Reals^2$, which allows us to consider momentum quantities as well.  Halvorson's argument suggests that he would advocate using the algebra $L^\infty(\mathbb{R}^2)$ in this case rather than our preferred $B(\mathbb{R}^2)$.}  The reason Halvorson gives is that we cannot distinguish by means of any observations or experiments between the situations where a particle's position lies in a set $A\subseteq\mathbb{R}$ and another set $B$ if $B\setminus A$ has Lebesgue measure zero.  But we contend that Halvorson's reasoning rests on a confusion between \emph{observables} and \emph{physical magnitudes} or \emph{quantities}.\footnote{For our purposes in this paper, we are taking the latter two notions of \emph{magnitude} and \emph{quantity} to be synonymous although we distinguish them from the notion of an \emph{observable}.}  If one takes the elements of an algebra to represent the physical magnitudes of a (classical or quantum) particle, then considerations of what we can come to know about those quantities do not obviously bear on their significance, nor on whether we have made an appropriate choice of algebra. 

However, despite this apparent difference in the algebra Halvorson uses for the classical case, Halvorson seems to agree with the standard view (implicitly accepted in this article) that in classical physics, there are physically significant states in which a classical particle has a determinate position.  It is a strange artifact that Halvorson must appeal to non-normal states on $L^\infty(\mathbb{R})$ because if he instead used a different classical algebra (e.g., $B(\mathbb{R})$, analogous to the situation we consider here), then determinate position states would appear already as normal states.  Halvorson only needs to advocate for using non-standard (non-normal) states because he has just advocated for the use of a non-standard algebra.  

In other words, there is a tension between Halvorson's justification for the algebra he uses and his justification for the collection of states he wants to deem physically possible.  On the one hand, he wants our algebra to take into account the finite precision of measurements of physical observables.  But on the other hand he imposes no such requirement on states, allowing non-normal states that have perfectly precise values for position.  If one fixes how one is using the mathematical apparatus to represent the physical system, then we do not see how one can simultaneously hold onto both views.  One can certainly have finite precision through and through and discuss only observables, or else one can allow for arbitrary precision and consider physical magnitudes.  But we see no justification for mixing and matching interpretations as Halvorson does.  We contend that Halvorson's perspective obscures the significance of the mathematical apparatus he uses---specifically the inherent connection between states and quantities.

We wish to pursue a strategy for relieving this tension.  We will drop the requirement of finite precision and explicitly use an algebra to represent physical magnitudes rather than observables.  This corresponds to our choice of the algebra of bounded measurable functions without identifying functions that agree almost everywhere.  These measurable functions can be understood as (perhaps idealized) physical quantities, even though the finite precision of our measurements may prevent us from distinguishing between them in measurement contexts.  Of course, to consider not just classical position magnitudes but also classical momentum magnitudes, one needs to shift attention from the position configuration space $\mathbb{R}$ to the phase space $\mathbb{R}^2$ of the particle, and hence use the algebra $B(\mathbb{R}^2)$, which brings us back to precisely the classical algebra considered above in Prop. \ref{prop:representation} and Cor. \ref{prop:normalrep} (see fn. \ref{fn:configurationphase}).

In this case, one sees immediately that normal states on $B(\mathbb{R}^2)$ are physically significant in the classical theory because they are the states that can be represented by Borel probability measures on the phase space $\mathbb{R}^2$.  This resolves one of the problems \citet{HalvorsonTeller} worries about---namely, whether it is possible to give an ignorance interpretation of probabilistic states.  The representation of normal states on $B(\mathbb{R}^2)$ as probability measures immediately fulfills the precondition for giving such an ignorance interpretation, where we are ignorant of which point in phase space can be used to represent the determinate state of the particle.  Moreover, notice that non-normal states on $B(\mathbb{R}^2)$ fail this test because they \emph{cannot} be represented by probability measures on the phase space $\mathbb{R}^2$.

Enough discussion of normality in classical theories---we now return to normality in quantum theories.  Recall that what we really care about is justifying the restriction to normal states on a quantum algebra in Cor. \ref{prop:normalrep}; we do so as follows.  If we accept that normal states on $B(\mathbb{R}^2)$ are the physically significant states of the classical particle, then a weakly continuous quantization map allows us to transfer the significance of normal states from the classical theory to normal states of the quantum theory.  Normal quantum states are ones whose composition with the quantization map are normal classical states.\footnote{This is an adaptation of Prop. 1 and Cor. 1 of \citet{Fe17b}.}  So on the basis of both a mathematical analogy and the physical significance with which we are endowing the algebraic structure of the quantized theory, one has reasons to use normal states in the quantum theory.  Thus, appeal to classical algebras and quantization maps provides at least one route to the justification for normal states that \citet{RuetscheNormal} claims is missing.  Appropriately continuous quantization maps help justify both the use of a particular algebra and its normal state space in one fell swoop.

We hope the preceeding paragraphs make clear that our algebraic ``imperialist'' stance differs significantly from that of Halvorson, who might more aptly be called an algebraic ``opportunist".  Halvorson goes back and forth between (often inconsistent) ways of interpreting operator algebras and their associated states.  We suggest a more careful and systematic approach to the way we employ algebraic tools, and we take this suggestion to be perhaps the most important moral of our discussion here.  When one does take this more cautious course, one finds a rather different perspective on regularity and normality.

\subsection{Quantization and regularity for approximate quantities}

We end this section by stating one further corollary to Prop. \ref{prop:representation} above, which we hope will clarify how the previous two propositions play out in practice.  One often quantizes a classical C*-algebra that does not even contain the classical Weyl unitaries.  In other words, one often quantizes a classical C*-algebra $\mathfrak{C}_0$ where $AP(\mathbb{R}^2)\nsubseteq\mathfrak{C}_0$.  One example is Berezin-Toeplitz quantization \citep[see][]{La98b}, which has $C_0(\mathbb{R}^2)$ as its domain.  The range of such a quantization map may be a C*-algebra $\mathfrak{A}$ that is not a W*-algebra.  However if one takes the quantization map $\mathcal{Q}_0:\mathfrak{C}_0\rightarrow\mathfrak{A}$ and finds its unique weak* continuous extension $\mathcal{Q}:\mathfrak{C}_0^{**}\rightarrow\mathfrak{A}^{**}$, then (as long as $\mathfrak{C}_0$ satisfies some technical constraints) one will have $AP(\mathbb{R}^2)\subseteq \mathfrak{C}_0^{**}$ and so the classical Weyl unitaries will at least appear in the domain of the extended quantization map $\mathcal{Q}$.  

In what follows, we will think of the classical algebra $\mathfrak{C}$ as such an extended algebra $\mathfrak{C}_0^{**}$, although one need not even assume $\mathfrak{C}$ is a W*-algebra to state the result.  Our extra assumption instead will be only that the quantum W*-algebra $\mathfrak{R}$ is the bidual of some C*-algebra $\mathfrak{A}$, i.e., $\mathfrak{R} = \mathfrak{A}^{**}$.  If one has prior reason to believe that $\mathfrak{A}$ is an algebra of physically significant magnitudes of the quantum particle and that magnitudes appearing in $\mathfrak{A}^{**}$ obtain their significance by weakly approximating those in $\mathfrak{A}$ (See \citet{Fe17c} for this perspective), then it is representations $\pi$ of $\mathfrak{A}$ that one should focus on.  From this perspective, one may wish to consider a representation of $\mathfrak{R}$ to be physically significant only insofar as it is obtained as the (unique) continuous extension $\tilde{\pi}$ to $\mathfrak{R} = \mathfrak{A}^{**}$ of a representation $\pi$ of $\mathfrak{A}$ .  Since any such representation $\tilde{\pi}$ is necessarily a w*-representation---without any further assumptions about ``preserving structure"---we have the following additional result.

\begin{cor}
\label{cor:c*}
Let $\mathfrak{C}$ be a C*-algebra with $AP(\mathbb{R}^2)\subseteq \mathfrak{C}\subseteq B(\mathbb{R}^{2})$.  Let $\mathcal{Q}:\mathfrak{C}\rightarrow\mathfrak{A}^{**}$ be a linear, *-preserving map into the bidual of a C*-algebra $\mathfrak{A}$.  Suppose $\mathcal{Q}$ is weakly continuous. Then the extension of any representation $\pi$ of $\mathfrak{A}$ to a w*-representation $\tilde{\pi}$ of $\mathfrak{A}^{**}$ is regular.
\end{cor}

\noindent Hence, if one obtains representations of a quantum algebra by extending representations of a C*-algebra constructed through a suitably continuous quantization map, then all such representations must be regular.  Thus, one can justify the use of w*-representations without appeal to their ``preserving structure'', and without invoking a restriction to normal states, by extending representations of C*-algebras to their biduals.  Then one arrives at the same conclusion, that physically significant representations are regular.

Now we have seen three senses in which proper attention to the physical significance of physical quantities in a C*-algebra, and how they are constructed from a corresponding classical theory, can be used to rule out non-regular representations even on a broadly algebraic imperialist approach.  This gives some reason---at least, provisional on the assumptions we attempted to motivate in this section---to focus on regular representations of an algebra representing the physical magnitudes of a quantum system.  Next, we recover some role for non-regular states in this approach, albeit a role different from that Halvorson proposes.

\section{Approximation of Non-Regular States}
\label{sec:approximation}

Having argued that the regular representations of a quantum algebra are physically privileged, we have in a sense come down on the side of \citet{Teller}, whom \citet{HalvorsonTeller} aims to refute.  Teller claims that quantum mechanics does \emph{not} allow for determinate position states, while Halvorson claims that quantum mechanics (understood through his algebraic perspective) \emph{does} allow for determinate position states.  We now seek a reconciliation between these two seemingly disparate views; we claim such a reconciliation can be found from the algebraic perspective provided in the preceeding section by revisiting Fell's theorem.  We will argue that even though the regular representations, and hence the states contained therein (called \emph{regular states}), are in a sense privileged,\footnote{The sense in which regular states are privileged is that one can justify restricting attention to them at least \emph{on the condition} one accepts the assumptions made in Prop. 1.} one can still show that states in non-regular representations (called \emph{non-regular states}) can be obtained as approximations or idealizations from these privileged states, according to a relevant notion of approximation.

Once we have the desired W*-algebra $\mathfrak{R}$ of Prop. \ref{prop:representation}, non-regular states may still appear as states on $\mathfrak{R}$.  However, in this case, non-regular states must, by Cor. \ref{prop:normalrep}, be \emph{non-normal} states.  So let us consider the relation of non-normal states to normal states on $\mathfrak{R}$, which we will assume are physically significant.  Notice that if we take the physical quantities in $\mathfrak{R}$ to be physically significant, then we have a physically salient notion of approximation in the weak* topology on the state space of $\mathfrak{R}$.  The relationship between normal states and non-normal states can be established with respect to this topology, as in the following result.

\begin{prop}
\label{prop:approx}
Let $\mathcal{S}(\mathfrak{R})$ and $\mathcal{S}_N(\mathfrak{R})$ denote the collections of states and normal states on a W*-algebra $\mathfrak{R}$, respectively.  Then $\mathcal{S}_N(\mathfrak{R})$ is dense in $\mathcal{S}(\mathfrak{R})$ in the weak* topology. 
\end{prop}

\noindent The intended interpretation of this result is that all non-normal states on $\mathfrak{R}$ can be approximated arbitrarily closely by normal states on $\mathfrak{R}$ according to a physically salient notion of approximation given by the weak* topology on the state space.  The result follows immediately from the version of Fell's theorem stated above.\footnote{Notice that Prop. \ref{prop:approx} concerns an arbitrary W*-algebra so, unlike the Halvorsonian use of Fell's theorem we outlined in \S\ref{sec:prelim}, our current setup does not depend on the peculiar fact that the Weyl algebra is simple.}  This shows a sense in which one can hold onto parts of both the standpoints of Teller (according to which regular states seem to be privileged) and Halvorson (according to which non-regular states are also somehow significant).  Teller's (\citeyear{Teller}) starting point seems to be correct that the regular states are privileged by virtue of being normal states on an appropriate algebra.  But on the other hand, one can still recover some physical significance for the non-regular or non-normal states, as \citet{HalvorsonTeller, HalvorsonBohr} desires, by showing that they can be obtained as approximations or idealizations from these privileged states.

It is important to note, however, that the physical significance the non-regular or non-normal states are endowed with on our interpretation is somewhat different from the significance \citet{HalvorsonBohr} attributes to them when he seems to count non-regular states as on a par with regular states.  On our interpretation, the non-regular states are obtained from the regular states as \emph{approximations} or \emph{idealizations}.  They are reached as limits, in a physically significant topology, from the physically significant states.

Moreover, we emphasize that this relationship is not generally symmetric: one could not just as well begin with the non-normal states and conclude that the normal states generically arise as idealizations from these.  In fact, there is a precise sense in which the non-normal states fail to approximate all of the normal states in the case of interest.  Consider a weakly continuous quantization into $\mathfrak{R} = \mathscr{B}(\mathscr{H})$, the bounded operators on a separable Hilbert space $\mathscr{H}$ (or in the notation of Cor. \ref{cor:c*} and Cor. \ref{cor:approx} below, we have $\mathfrak{A} = \mathscr{K}(\mathscr{H})$, the compact operators).\footnote{This choice of algebra could arise, e.g., from Berezin-Toeplitz quantization \citep[see][]{La98b,BeCo86}.}  Let $\mathcal{S}_S(\mathfrak{R})$ be the convex subset of $\mathcal{S}(\mathfrak{R})$ generated by the pure \emph{non}-normal states on $\mathfrak{R}$; the subscript $S$ here denotes that these are the \emph{singular} states \citep[pp. 722-723]{KaRi97}.\footnote{One might be tempted to consider the collection of all non-normal states, but this will not do because one can too easily find non-normal states by taking mixtures of normal states with singular states.  It then becomes trivial to approximate any normal state by an appropriate mixture simply by letting the singular component tend to zero.  This is clearly not what we are looking for; we would like ask whether normal states can be approximated by states that have no normal component, which are precisely the singular states.}  For example, if one takes $\mathscr{H}$ to be the Hilbert space carrying the Schr\"{o}dinger representation of the Weyl algebra, then the determinate position and determinate momentum states on the Weyl algebra can be extended to states on $\mathscr{B}(\mathscr{H})$, and these extensions are singular.  In this case, $\mathcal{S}_S(\mathfrak{R})$ \emph{fails to be weak* dense in} $\mathcal{S}(\mathfrak{R})$, which shows a sense in which regular states cannot in general be obtained as limiting approximations from non-regular states.\footnote{See \ref{app:proofs} for a sketch of a proof.}  (That non-regular states are dense in the regular states on the Weyl algebra is a peculiarity, arising because the Weyl algebra is simple.) 

Our interpretation also differs from a point Halvorson makes in the earlier paper (\citeyear{HalvorsonTeller}), in which he argues for the significance of non-normal states.  There, Halvorson claims that non-normal states are needed to serve as the basis for an ignorance interpretation of probabilities in quantum mechanics so that quantum states can be understood as measures of ignorance over determinate states.  From the current perspective, it would be very strange if we were forced to think of quantum states as expressing our ignorance about which idealized or approximate state a system is in when the system is actually in a non-idealized state.

One remark is in order to clarify the status of the approximation result in Prop. \ref{prop:approx}.  Why does the weak* topology have the appropriate physical significance here?  This is especially important given that we questioned precisely this assumption in the interpretation of Fell's theorem outlined in \S\ref{sec:prelim}.  Recall that the weak* topology on $\mathcal{S}(\mathfrak{R})$ is the notion of approximation for states induced by expectation values on quantities in $\mathfrak{R}$.  It follows that if $\mathfrak{R}$ does indeed consist of precisely the physically significant quantities of the system, then the weak* topology on $\mathcal{S}(\mathfrak{R})$ is likewise physically significant.  The difference between our situation now and that described in \S\ref{sec:prelim} is that we now suppose we have managed to fix, whether by the constraints of a quantization map or by some other means, an appropriate W*-algebra of physical magnitudes of our system.  This will not be the Weyl algebra if the constraints discussed above are satisfied, so we do not arrive back at the Halvorsonian argument outlined in \S\ref{sec:prelim}.

Finally, we state a further result by assuming we are in the setup of Cor. \ref{cor:c*}, where the W*-algebra $\mathfrak{R}$ is the bidual of some C*-algebra $\mathfrak{A}$, i.e., $\mathfrak{R} = \mathfrak{A}^{**}$.  Recall that this situation characterizes many quantization procedures actually in use.  In this case, Prop. \ref{prop:approx} has the following immediate corollary:
\begin{cor}
\label{cor:approx}
Let $\mathcal{S}(\mathfrak{A})$ and $\mathcal{S}(\mathfrak{A}^{**})$ denote the collections of states on a C*-algebra $\mathfrak{A}$ and the W*-algebra $\mathfrak{A}^{**}$.  Then $\mathcal{S}(\mathfrak{A})$ is dense in $\mathcal{S}(\mathfrak{A}^{**})$ in the weak* topology.
\end{cor}
\noindent In this case, the normal states can be understood as states on an original C*-algebra $\mathfrak{A}$ of physically significant quantities, while the non-normal states only appear on the extended algebra $\mathfrak{A}^{**}$ containing approximate or idealized quantities.  Thus, the intended interpretation of Cor. \ref{cor:approx} is again that non-normal states can be approximated arbitrarily closely by normal states with respect to a physically significant notion of approximation.  We maintain, however, that this interpretation is compatible with understanding the regular (and normal) states to be privileged, and in this sense the non-normal states might be used as idealizations.  Thus, our conclusions differ significantly from Halvorson's.

We emphasize that the results of this section are just a further application of Fell's theorem.  Now, however, we have reason to think that some states have \emph{particular} physical significance by using the information encoded in a classical theory and quantization map.

Thus, we provide a justification of both the regularity condition and the approximation of non-regular states, and this justification goes beyond mere instrumental value or technical convenience.  Moreover, this justification does not beg the question against Halvorson because it comes from an external source---the classical theory being quantized.  Of course, one is still free to question our interpretation of the information from the classical theory; but even if one does, we have still made progress by pushing the question back to the more familiar context of classical physics.  What we have shown is that if one does accept our interpretation of the classical theories being quantized, then the regularity assumption can be seen to follow from natural constraints on a quantization map relating the magnitudes of our classical and quantum theories.  If one wants to appeal to non-regular states, we have at the very least shown precisely what constraints one must give up to do so.

\section{Conclusion}
\label{sec:conclusion}

In this paper, we have attempted to justify the regularity condition by putting constraints on the construction of quantum theories in the process of quantization.  If correct, this provides reason to rule out the position and momentum representations \citet{HalvorsonBohr} uses to demonstrate the existence of states with determinate position or momentum values.  No determinate position or momentum states can be found in regular representations.  However, we went on to propose a reconciliation with Halvorson by establishing that these non-regular determinate position and momentum states can still be obtained as approximations or idealizations.  This gives rise to a different interpretation than Halvorson's, but might make our conclusion more palatable to advocates of non-regular representations.

Perhaps most important for future philosophical work is the algebraic perspective we have tried to exemplify in this paper.  One might get the impression from the literature---e.g., the description of algebraic imperialism found in \citet{Arageorgis} or \citet{Ruetsche}---that there is only one way to use algebraic methods to interpret quantum theories.  We hope to have demonstrated here that this is emphatically \emph{not} the case.  We believe some of the most pressing matters in the interpretation of quantum theories appear by looking at context-specific questions concerning algebraic quantum theories, which go beyond merely choosing whether to use operator algebras or Hilbert spaces.  We believe these issues are best approached not by considering the status of mathematical tools abstractly, but rather \emph{how mathematical tools are used to represent particular physical systems}.

We worry that the alternative approach---in which one argues at a high level of generality---can be misleading.  For example, \citet{Ea17,Ea17a} and \citet{ArEaRu17} consider additivity conditions for quantum states as probability assignments and the possibility of subjective interpretations of these probability assignments.  We are friendly to many these authors' conclusions, but we worry about their methods of argument.  In some instances, these authors provide general arguments for countable additivity, or equivalently normal states, without any attention to the physical systems these tools might be used to represent (see, e.g., the ``transcendental argument'' in \citet[pp. 26-28]{ArEaRu17}).  From the perspective of this paper, which agrees with the earlier claims of \citet{RuetscheNormal}, it is absolutely essential to first understand or interpret the physical quantities or propositions that one is assigning expectation values or probabilities to \emph{before} one can decide whether states or probability measures are physical or how to interpret them.  Indeed, we have demonstrated that which states one considers physical may depend in subtle ways on which algebra of physical magnitudes one uses---even in the very simplest quantum theories considered here.

We hope to have illustrated in this paper, however, that it \emph{is} possible for philosophers to engage in the detailed and context-specific interpretive work required to understand physical magnitudes, physical states, and the relations between them in classical and quantum theories.  Since we have only had the opportunity here to consider very simple (some might even say trivial!) systems, we hope others will join us in extending this algebraic perspective to philosophical investigations of further issues in quantum theories.

\section*{Acknowledgments}
We are grateful to Hans Halvorson, David Malament, and Laura Ruetsche for helpful conversations related to this material.

\appendix

\section{Algebraic Background}
\label{app:background}

A \emph{C*-algebra} $\mathfrak{A}$ is an involutive, associative, complete normed algebra satisfying
\[\norm{A^*A} = \norm{A}^2\]
for all $A\in\mathfrak{A}$.  The dual space $\mathfrak{A}^*$ is the collection of bounded linear functionals on $\mathfrak{A}$. A \emph{state} $\omega$ on $\mathfrak{A}$ is a positive ($\omega(A^*A)\geq 0$ for all $A\in\mathfrak{A}$) and normalized ($\norm{\omega} = 1$) linear functional on $\mathfrak{A}$.  The collection of all states is called the \emph{state space} of $\mathfrak{A}$, denoted $\mathcal{S}(\mathfrak{A})$.

A \emph{representation} $(\pi,\mathscr{H})$ of $\mathfrak{A}$ consists in a Hilbert space $\mathscr{H}$ and a *-homomorphism $\pi:\mathfrak{A}\rightarrow\mathcal{B}(\mathscr{H})$ into the bounded linear operators on $\mathscr{H}$.  One can define a number of useful notions with the tools of representations.  The weak operator topology on a representation is characterized by the following condition of convergence.  A net $A_\alpha\in\mathcal{B}(\mathscr{H})$ converges to $A\in\mathcal{B}(\mathscr{H})$ if for all vectors $\varphi,\psi\in\mathscr{H}$, $\inner{\varphi}{A_\alpha\psi}\rightarrow\inner{\varphi}{A\psi}$.  A state $\omega\in\mathcal{S}(\mathfrak{A})$ has a \emph{density operator representative} in $(\pi,\mathscr{H})$ if there is a density operator $\rho_\omega$ on $\mathscr{H}$ such that
\[\omega(A) = Tr(\rho_\omega A)\]
for all $A\in\mathfrak{A}$.

Given any state $\omega\in\mathcal{S}(\mathfrak{A})$, the \emph{GNS construction} yields a canonical representation $(\pi_\omega,\mathscr{H}_\omega)$ of $\mathfrak{A}$ containing a unit vector $\Omega_\omega\in\mathscr{H}_\omega$ such that
\[\omega(A) = \inner{\Omega_\omega}{\pi_\omega(A)\Omega_\omega}\]
for all $A\in\mathfrak{A}$.  The representation $\pi_\omega$ is unique up to unitary equivalence and called the \emph{GNS representation of $\mathfrak{A}$ for $\omega$}.  A state $\omega$ clearly always has a density operator representative in its own GNS representation.  The \emph{universal representation} of $\mathfrak{A}$ is the representation $(\pi_U,\mathscr{H}_U)$ given by
\[\pi_U(A) = \bigoplus_{\omega\in\mathcal{S}(\mathfrak{A})}\pi_\omega(A)\]
on
\[\mathscr{H}_U = \bigoplus_{\omega\in\mathcal{S}(\mathfrak{A})}\mathscr{H}_\omega\]
where $(\pi_\omega,\mathscr{H}_\omega)$ is the GNS representation of $\mathfrak{A}$ for $\omega$.  Every state on $\mathfrak{A}$ has a density operator representative in the universal representation.

One can also define useful notions in the absence of a representation.  The \emph{weak topology} on $\mathfrak{A}$ is given by the following condition for convergence: a net $A_\alpha\in\mathfrak{A}$ converges to $A\in\mathfrak{A}$ if for all $\omega\in\mathfrak{A}^*$, $\omega(A_\alpha)\rightarrow\omega(A)$.  Although a C*-algebra is required to be complete in its norm, it is typically not complete in the weak topology.  However, one can always construct the weak completion of a C*-algebra $\mathfrak{A}$ by taking its bidual $\mathfrak{A}^{**}$, the collection of bounded linear functionals on $\mathfrak{A}^*$.  The bidual $\mathfrak{A}^{**}$ is a C*-algebra that is furthermore the dual space to a Banach space.  This further condition makes $\mathfrak{A}^{**}$ a W*-algebra.  A \emph{W*-algebra} $\mathfrak{R}$ is a C*-algebra that is the dual space to a Banach space $\mathfrak{R}_*$, i.e., $\mathfrak{R} = (\mathfrak{R}_*)^*$.  The \emph{weak* topology} on the dual space $V^*$ to a vector space $V$ is characterized by the following condition for convergence: a net $v_\alpha\in V$ convergences to $v\in V$ if for all $u\in V^*$, $v_\alpha(u)\rightarrow v(u)$.  Thus, there is a weak* topology on both a W*-algebra $\mathfrak{R}$ and the dual space $\mathfrak{A}^*$ to any C*-algebra $\mathfrak{A}$, and furthermore, there is a weak* topology on $\mathcal{S}(\mathfrak{A})$, understood as a subspace of $\mathfrak{A}^*$.  With these definitions, a W*-algebra is complete in its weak* topology.  Finally, a state $\omega$ on a W*-algebra $\mathfrak{R}$ is a \emph{normal} state if there is an element $\hat{\omega}\in \mathfrak{R}_*$ such that $\omega(A) = A(\hat{\omega})$ for all $A\in\mathfrak{R}$.  The normal states are precisely the weak* continuous states on $\mathfrak{R}$.

\section{Proofs of Propositions}
\label{app:proofs}

\begin{prop*}{1}
Let $\mathfrak{C}$ be a C*-algebra with $AP(\mathbb{R}^2)\subseteq \mathfrak{C}\subseteq B(\mathbb{R}^{2})$.  Let $\mathcal{Q}:\mathfrak{C}\rightarrow\mathfrak{R}$ be a linear, *-preserving map into a W*-algebra $\mathfrak{R}$.  Suppose further that $\mathcal{Q}$ is weakly continuous.    Then every w*-representation of $\mathfrak{R}$ is regular.
\end{prop*}

\begin{proof}
We know from the continuity of $\mathcal{Q}$ that $a\mapsto\mathcal{Q}(w_{a,0})$ and $b\mapsto\mathcal{Q}(w_{0,b})$ are weak* continuous one-parameter families in $\mathfrak{R}$ because $a\mapsto w_{a,0}$ and $b\mapsto w_{0,b}$ are weak* continuous one-parameter families on $B(\mathbb{R}^2)$.  Thus, since $\pi$ is continuous in the weak* and weak operator topologies, it follows that $a\mapsto\pi\circ\mathcal{Q}(w_{a,0})$ and $b\mapsto\pi\circ\mathcal{Q}(w_{0,b})$ are weak operator continuous one-parameter families.
\end{proof}

\noindent Cor. \ref{prop:normalrep} follows immediately from Prop. \ref{prop:representation} and the following lemma, which summarizes the discussion on p. 41 of \citet{Sa71} (preceding Thm. 1.16.8):

\begin{lem}
\label{lem:GNS}
Let $\omega$ be a normal state on a W*-algebra $\mathfrak{R}$.   The GNS representation $(\pi_\omega,\mathcal{H}_\omega)$ for $\omega$ is a w*-representation.
\end{lem}

\noindent Cor. \ref{cor:c*} then follows immediately from Prop. \ref{prop:representation} and the fact that $\mathfrak{A}^{**}$ is a W*-algebra.

\begin{prop*}{4}
Let $\mathcal{S}(\mathfrak{R})$ and $\mathcal{S}_N(\mathfrak{R})$ denote the collections of states and normal states on a W*-algebra $\mathfrak{R}$, respectively.  Then $\mathcal{S}_N(\mathfrak{R})$ is dense in $\mathcal{S}(\mathfrak{R})$ in the weak* topology. 
\end{prop*}

\begin{proof}
By Thm. 1.16.7 of \citet[][p. 42]{Sa71}, the universal normal representation $\pi_N := \bigoplus_{\omega\in\mathcal{S}_N(\mathfrak{R})}\pi_\omega$ is a faithful representation of $\mathfrak{R}$.  The weak* continuity of this representation implies that every vector state in $\pi_N$ is normal, which shows that the vector states in $\pi_N$ are all and only the normal states on $\mathfrak{R}$.  Likewise, the universal representation $\pi_U:= \bigoplus_{\omega\in\mathcal{S}(\mathfrak{R})}\pi_\omega$ is also a faithful representation of $\mathfrak{R}$ \citep[See, e.g., Thm. 1.16.6, p. 41]{Sa71}.  And every state on $\mathfrak{R}$ is a vector state in $\pi_U$.  Since the representations $\pi_N$ and $\pi_U$ agree on their kernel, it follows from Thm. 1.2 of \citet{Fe60} (in the direction (i)$\Rightarrow$(iv)) that every vector state on $\pi_U$ is a weak*-limit of finite linear combinations of vector states on $\pi_N$.  This implies that every state in $\mathcal{S}(\mathfrak{R})$ can is a  weak* limit of states in $\mathcal{S}_N(\mathfrak{R})$, which completes the proof.
\end{proof}

\noindent Cor. \ref{cor:approx} follows immediately from Prop. \ref{prop:approx} and the fact that $\mathfrak{A}^{**}$ is a W*-algebra whose normal state space is the state space of $\mathfrak{A}$.

\begin{prop*}{6}
Let $\mathfrak{R}$ be the W*-algebra $\mathscr{B}(\mathscr{H})$ of bounded operators on a separable Hilbert space $\mathscr{H}$.  Let $\mathcal{S}(\mathfrak{R})$ and $\mathcal{S}_S(\mathfrak{R})$ denote the collections of states and singular states on $\mathfrak{R}$, respectively.  Then $\mathcal{S}_S(\mathfrak{R})$ fails to be dense in $\mathcal{S}(\mathfrak{R})$ in the weak* topology.
\end{prop*}

\begin{proof}
We merely sketch an argument here.  Begin by noting that every singular state on $\mathfrak{R}$ annihilates the compact operators \citep[Cor. 10.4.4, p. 750]{KaRi97}.  One can then either show directly that this means no (normal) state that is non-zero on the compact operators can be obtained as a weak* limit of singular states, or else equivalently note that this implies \citep[Thm. 10.4.6, p. 751]{KaRi97} that any representation of $\mathfrak{R}$ with only singular states in its folium has a kernel that contains the compact operators whereas the universal representation has trivial kernel, and then apply Theorem 1.2 of \citet{Fe60} in the converse of the direction cited above (i.e., in the direction (iv)$\Rightarrow$(i)).
\end{proof}

\singlespacing

\bibliographystyle{elsarticle-harv}
\bibliography{regularity}
\end{document}